\documentclass[runningheads,a4paper]{llncs}

\usepackage{latexsym,amssymb,amsmath}
\usepackage{epsf}
\usepackage{graphicx}
\usepackage{subfigure}

\begin{document}

% git push && ssh g3 cd /tmp/iclp08'&&'git push

\title{Better Termination for Prolog with Constraints}

\author{Markus Triska\inst{1} \and Ulrich Neumerkel\inst{1} \and Jan Wielemaker\inst{2}}
\institute{Technische Universit\"{a}t Wien, Austria\\
	   \email{\{triska,ulrich\}@complang.tuwien.ac.at}\and
           Universiteit van Amsterdam, The Netherlands\\
	   \email{J.Wielemaker@uva.nl}}

\maketitle
\setcounter{page}{90}

\begin{abstract}

Termination properties of actual Prolog systems with constraints are
fragile and difficult to analyse.  The lack of the occurs-check, moded
and overloaded arithmetical evaluation via \verb+is/2+ and the occasional
nontermination of finite domain constraints are all sources for
invalidating termination results obtained by current termination
analysers that rely on idealized assumptions.  In this paper, we
present solutions to address these problems on the level of the
underlying Prolog system.  Improved unification modes meet the
requirements of norm based analysers by offering dynamic occurs-check
detection.  A generalized finite domain solver overcomes the
shortcomings of conventional arithmetic without significant runtime
overhead.  The solver offers unbounded domains, yet propagation always
terminates.  Our work improves Prolog's termination and makes Prolog a
more reliable target for termination and type analysis.  It is part of
SWI-Prolog since version 5.6.50.

\end{abstract}

%%%%%%%%%%%%%%%%%%%%%%%%%%%%%%%%%%%%%%%%%%%%%%%%%%%%%%%%%%%%%%%%
\section{Introduction}

Termination plays a central role in Prolog programs.  Prolog's complex
control mechanism often taxes a programmer's intuition about
termination.  Tools to support both experts and beginners are
therefore highly valuable and the development of such systems has
received considerable attention \cite{arithan,normbased,cti}.  One of the authors
was particularly interested in developing termination tools for
supporting beginners within the learning environment GUPU
\cite{gupu2002}.  In a collaborative effort, the termination
inference system cTI \cite{cti} was developed that featured not only a
web interface but was designed to specifically meet the incremental
demands for an on-the-fly analyser by employing a strict bottom up
approach.

Much to our chagrin, the resulting system soon showed the limitations
of current approaches for our original goals.  cTI worked
quite impressively for current benchmarks but did not reflect the
entire spectrum of termination properties of actual Prolog
implementations.  cTI---like most other norm based approaches
\cite{normbased}--- was founded on some assumptions that are not true
for existing Prolog systems.  As a consequence, the termination
conditions inferred with cTI are not literally applicable to the target
system---at that time SICStus Prolog.  We note that these problems do
not show in existing termination benchmarks, but are frequently
occurring in the incorrect programs beginners write.  The source of
the problem is the lack of the occurs-check in existing Prolog
implementations giving way to rational trees that can no longer be
mapped onto the integers.  While there are approaches to determine
occurs-check freeness statically \cite{occurcheck}, as well as finite
trees \cite{finitetree} we finally chose to go for the
maximum which is performing the occurs-check dynamically.

With the addition of constraints to Prolog's core language, new
sources of unforeseen nontermination opened, further complicating
procedural reasoning.  The traditional \verb+is/2+ predicate with its
overloaded semantics posed even more problems.  To meet all these needs
we implemented a new version of a generalized finite domain solver.
This library subsumes the functionality of integer arithmetic and
constraint programming, combines their strengths, and terminates
always, permitting better termination results.

\paragraph{Content.} We first describe our new approach to the old
occurs-check problem and then discuss our improvement to clpfd to
subsume \verb+is/2+-functionality.  Finally we present our new always
terminating implementation of clpfd.

\section{Occurs-check}

Most existing Prolog implementations use rational tree unification
\cite{colme1982} to avoid overheads caused by the occurs-check of
finite tree unification.  While rational trees are an interesting
domain in their own right, they are often an indication for
programming errors.  For beginners, it is very common to accidentally
confuse assignment and unification.  Goals like \verb+Xs = [X|Xs]+ are
often written with the intention to add to the list \verb+Xs+ an
element.  Also misunderstandings concerning the scoping of variables
lead to infinite terms.  Exactly such cases are not covered by
existing norm based approaches that assume the finiteness of terms.

We added two new standard conforming unification modes that prevent
the creation of infinite terms.  Apart from traditional occurs-check
that fails silently, a new mode was added to better
localize attempts to create infinite terms.
By issuing \verb+?- set_prolog_flag(occurs_check,error).+ at runtime all attempts
to create infinite terms are detected and an error is
issued.  In this manner all programs are identified that create
infinite terms.  Also, most programs subject to
occurs-check (STO) are detected, that are ruled out by the ISO standard
\cite{iso1995}.

Our implementation tries to avoid the costly occurs-check scan for the
most frequent cases of passing variables.  Current Prolog
implementations allocate variables that do not occur within a
structure in a separate storage area, mostly known as the goal or
environment stack.  Those variables are unified in constant time with
structured terms, as the cannot be the subterm of a structure.  In
this manner most uses of difference lists and differences with other
data structures do not require the occurs-check.  The actual testing
can be further reduced taking into account that Prolog compilers emit
specialised unification instructions where possible, based on its
knowledge about the arguments involved in the unification.  Only the
cases of instructions of general unification are subject to occurs
check. All other cases do not involve any overhead.  As of version
5.7, all overheads for handling the list differences of DCGs are
completely removed for an initial goal \verb+phrase/2+.  For
\verb+phrase/3+ there is a single occurs-check for each solution
found.

\section{Overcoming {\tt is/2}}

Using \verb+is/2+ in pure programs has many disadvantages.  For one, \verb+is/2+
works only for certain restricted modes thereby limiting the
relational view of a predicate.  This relational view permits to test
programs more extensively---testing them with generalized modes.  Even
if those generalized modes are not used in the final application, they
help to detect otherwise undiscovered problems.  Consider for
example McCarthy's ``mysterious'' 91-function.  With the following
query we search for results different to 91.

\begin{verbatim}
mc_carthy_91(X, Y) :-
   X #> 100, Y #= X - 10.
mc_carthy_91(X, Y) :-
   X #=< 100, Z #= X + 11,
   mc_carthy_91(Z, Z1),
   mc_carthy_91(Z1, Y).

?- Y #\= 91, mc_carthy_91(X, Y).
Y in 92..sup,
-10+X#=Y,
X in 102..sup ;
(looping)
\end{verbatim}

Attempts to emulate with \verb+is/2+ different modes require the explicit
usage of \verb+var/1+ and \verb+nonvar/1+, two built-ins that lead frequently to
errors due to forgotten modes.

The overloading of integer and floating-point arithmetic is
another source of frequent errors with \verb+is/2+.  An accidentally
introduced float might lead to unexpected failures.  Modeling without
knowing whether or not a variable is a float is not reliably possible,
thereby weakening termination analysis \cite{arithan}.

For these reasons we propose to use in place of \verb+is/2+ the
corresponding \verb+#=/2+ of clpfd and the corresponding comparison
relations.  To make this shift more practical we removed the common
limits of \verb+#=/2+ to small integers and improved execution for
such simple moded cases.  While using \verb+#=/2+ in place of
\verb+is/2+ incurred overheads greater than two orders of magnitude for
small loops, our improved implementation is only about 30\% slower
than naive \verb+is/2+.  In this manner, we obtain predicates that are
simpler to type and that are not moded.

The original version of \verb+factorial/2+ is not tail recursive due to the
modedness of \verb+is/2+.  The space for allocating the environments in the
original version is traded for allocating constraints on the global
stack.  \verb+factorial/2+ now terminates if
either the first argument is finite, or the second argument is finite and not equal zero.

\begin{verbatim}
factorial(0, 1).
factorial(N, F) :-
   N #> 0,
   F #= F0*N,
   N1 #= N - 1,
   factorial(N1, F0).

?- Y in 1..5,  factorial(X,Y).
Y = 1,
X = 0 ;
Y = 1,
X = 1 ;
Y = 2,
X = 2 ;
false.
\end{verbatim}

\section{Terminating constraints}

Current implementations of finite domain constraints are optimized for
the traditional usage pattern of constraint satisfaction.  First,
variables get their associated domains, then the constraints between
variables are posted, and finally labeling searches for actual
solutions.  In current implementations, the declaration of a
variable's domain is just a simple goal.  (Original systems
required a static declaration.)  The extension from this limited view
toward a general constraint systems over integers, a kind of CLP(Z), is
straightforward.

By accepting variables without a finite domain, we open the
door to nonterminating constraint propagation.  Consider the query
\verb+?- X#>Y, Y#>X, X#>0.+ Existing constraint solvers will try to
reduce the domains until the maximal domain value is encountered, then
failing or yielding a representation error.  We therefore consider
this case the same as genuine nontermination. Note that
nontermination does not only occur due to posting a constraint but also
may happen during labeling.

\verb+?- X#>Y, Y#>X, X#>B*Y, B in -1..0, labeling([],[B])+

Termination within constraint propagation is ensured by propagating
domain changes in infinite domains only once.  At the price of
weakening consistency we can now guarantee that clpfd and all
unifications with constrained variables terminate.

\subsection{Observing termination}

The notion of termination and nontermination are idealizations of
actual observable behavior that lead to seemingly paradoxical situations.
The query \verb+?- X#>X*X.+ terminates rapidly in SICStus 3 with a representation error.
Still, we consider this a case of non-termination.
For \verb+?- abs(X)#<7^7^7,X#>Y,Y#>X.+ in SWI, termination is not observable
within reasonable time.  However, we consider this case terminating.

Another rather unintuitive consequence concerns the termination
property of the entire program. While our improvement guarantees
termination for unification and all clpfd-goals, and therefore might
improve termination of the entire program, there are cases where a
stronger propagation that does not terminate in the general case will
nevertheless result in better termination of the entire program.  This
may happen, if the stronger propagation results in failure preventing
an infinite loop, while terminating propagation yields
inconsistency.

\subsection{Ad hoc termination proofs}

With an always terminating clpfd, we are able to perform some simple
forms of termination testing when using labeling.  One frequent
problem with larger constraint problems concerns the time span to wait
for the first solution.  Quite often labeling is considered to be
inefficient, when in reality the actual predicate definition that
posts the constraints does not terminate.  To avoid this situation we
separate the actual relation from labeling.  In place of the original
predicate \verb+p/+$n$ we define a new relation \verb+p_/+$n+1$
(``core relation'') that
contains an additional argument for the list of variables to be
labeled.  Consider for example as original query
\verb+?- queens(Ds).+ describing solutions for a given fixed length of
\verb+Ds+.  This query is now formulated as
\verb+?- queens_(Ds,Zs), labeling([],Zs).+ Suppose now that the answer
does not appear immediately.  Should we wait for an answer?  What, if
the query does not terminate?  To better understand the termination
properties involved we can consider the following query.  If \verb+?- queens_(Ds,Zs), false.+
terminates (by observation), we know also that the query followed by labeling will
terminate, since in our implementation \verb+labeling/2+ is guaranteed to terminate.  We
thus obtain a proof for termination by observing the termination of
another related predicate.
In systems without our favorable termination property, a terminating
\verb+?- queens_(Ds,Zs), false.+ does not constitute a termination proof of the goal followed by a search with \verb+labeling/2+.

\subsection{Black-box testing}

While developing and testing library(clpfd), it soon became evident
that manual testing and testing with given applications is not
sufficient.
  We noted as one of the most prominent coding errors the omission
of certain rare cases of instantiations.  The current implementation
in Prolog based on hProlog-style attributed variables \cite{attvar} does not guarantee any
properties concerning the correctness of the implementation.  The
concerns consistency and correctness must be dealt with on the same
level - thereby increasing the chance for errors.  As one of the
authors experienced similar problems with other constraint
implementations prior to SWI, it was evident that a more systematic
approach was needed.  Existing approaches to testing and specifying
finite domain constraints \cite{apt} were also not very attractive, as
they require considerable effort for specifying the actual propagation
mechanism.  Such complex specifications may again be a further source
of errors.  We therefore focused on testing with strictly minimal
information - thereby minimizing demotivating cases of false alarms.

We concentrated on testing a fixed set of algebraic properties for
small finite domains.  So far, all encountered correctness errors
could be shown to violate those properties.  We illustrate our
approach with an error located in this manner (i3a\#98).  The query
\verb+?- X in 0..2, 0/X#=0.+ should succeed, but failed.  Even to the
experienced constraint programmer it is not obvious by naively
inspecting this query what the correct result should be.  The bug was
located automatically by detecting a difference between the following
two queries:

\begin{verbatim}
  ?-        X in 0..2, 0/X #= 0, X = 1.
  ?- X = 1, X in 0..2, 0/X #= 0.
\end{verbatim}

The first query failed, the second succeeded.  Evidently,
there must be at least one error---either in the first or second
query, or in both.
Most errors found are related to the implementation of nonlinear
constraints like general multiplication.  Also, sharing of variables
was a frequent cause for errors. In total, more than 30 errors of this kind
were found by systematically exploring a tiny slice of all possible
formulae.

For efficient testing, (rapid) termination of clpfd's propagation is
indispensable.  This permits to test many queries simultaneously.  On
systems with nonterminating propagation, we would have to rely on
timeout mechanisms to interrupt certain queries that cannot be tested
in this way.

\subsection{Related work}

SICStus Prolog \cite{sicstus1997} was the first system to generalize
finite domain constraints without sacrificing correctness.  It uses
small integers for domains but signals domain overflows as
representation errors and not as silent failures.

Built-in support for the occurs-check has been implemented with
similar techniques  in Sepia Prolog
\cite{sepia} and its successor Eclipse Prolog \cite{eclipse}.

%%%%%%%%%%%%%%%%%%%%%%%%%%%%%%%%%%%%%%%%%%%%%%%%%%%%%%%%%%%%%%%%
\section{Conclusions}

The presented improvements constitute a more solid target for
termination analysis than prior implementations.  We hope that they
will lead to the development of more powerful analysers.

\paragraph{Acknowledgements.}
We would like to thank Philipp Kolmann of ZID for his generous support
on the Condor based \cite{condor} WINZIG-grid \cite{kolmann2005} at TU
Wien that we use to perform large scale tests.

\newpage
\bibliographystyle{plain}

\begin{thebibliography}{55}

\bibitem{apt}K.R.~Apt, P.~Zoeteweij.  An Analysis of Arithmetic
  Constraints on Integer Intervals, Constraints 12(4). 2007.

\bibitem{finitetree}R.~Bagnara, E.~Zaffanella, R.~Gori, P.~Hill.
  Boolean functions for finite-tree dependencies. LPAR 2001.

\bibitem{sicstus1997}M.~Carlsson, G.~Ottosson, B.~Carlson. An
  Open-Ended Finite Domain Constraint Solver. PLILP. 1997.

\bibitem{normbased}M.~Codish, C.~Taboch. A Semantic Basis for the
  Termination Analysis of Logic Programs. JLP. 41(1), 1999.

\bibitem{colme1982}A.~Colmerauer.  Prolog and Infinite Trees.
  Logic Programming, K.L.~Clark, S.-A.~T{\"a}rnlund (eds), 1982.

\bibitem{occurcheck}L.~Crnogorac, A.~Kelly, H.~S{\o}ndergaard. A
  comparison of three occur-check analysers. SAS 1996.

\bibitem{attvar}B.~Demoen. Dynamic attributes, their hProlog
  implementation, and a first evaluation, TR CW 350, Leuven, 2002.

\bibitem{arithan}N.~Dershowitz, N.~Lindenstrauss, Y.~Sagiv,
  A.~Serebrenik. Automatic Termination Analysis of Programs Containing
  Arithmetic Predicates. Electr. Notes TCI 30(1), 1999.

\bibitem{yap}A.~Faustino da Silva, V.~S.~Costa.  The Design and
  Implementation of the {YAP} Compiler, ICLP 2006.

\bibitem{iso1995}ISO/IEC 13211-1 Programming languages - Prolog - Part
  1: General core. 1995.  (7.3 Unification)

\bibitem{kolmann2005}Ph.~Kolmann. University Campus Grid
  Computing. Diploma thesis, TU Wien, 2005.

\bibitem{cti}F.~Mesnard, U.~Neumerkel.  Applying Static Analysis
  Techniques for Inferring Termination Conditions of Logic
  Programs. SAS 2001.

\bibitem{sepia}M.~Meier.  SEPIA: A Basis for Prolog Extensions.  LPAR
  1992.

\bibitem{gupu2002}U.~Neumerkel, St.~Kral.  Declarative program
  development in Prolog with GUPU.  WLPE.  2002.

\bibitem{condor}D.~Thain, T.~Tannenbaum, M.~Livny.  Condor and the
  Grid.  In F.~Berman, G.~Fox, A.~Hey (ed.), Grid Computing: Making
  the Global Infrastructure a Reality.  Wiley, 2003.

\bibitem{eclipse}M.~Wallace, St.~Novello, J.~Schimpf. ECLiPSe: A
  Platform for Constraint Logic Programming, ICL systems yournal 1997.

\bibitem{swi}J.~Wielemaker. An Overview of the {SWI-Prolog}
  Programming Environment, WLPE 2003.

\end{thebibliography}

\end{document}